# Valid for Much of "Realistic" Fields: A Non-Generational Conjecture for Deriving All First-Class Constraints At Once

K. Rasem Qandalji
*Amer Institute*
*P.O. Box 1386, Sweileh, 11910*
*JORDAN*
*E-mail*: qandalji@hotmail.com

**ABSTRACT**

We propose a single-step non-generational conjecture for derivation of all first class constraints, (involving, only, variables compatible with canonical Poisson brackets), of a realistic gauge (singular) field theory. We verify our conjecture for the free electromagnetic field, the Yang-Mills fields in interaction with spinor and scalar fields, and we also verify our conjecture in the case of gravitational field. We show that the first class constraints, which were reached at using the standard Dirac's multi-generational algorithm, will be reproduced using the proposed conjecture. We make no claim that this conjecture is valid for all "mathematically" plausible Lagrangians; but, nevertheless, the examples we consider here show that this conjecture is valid for a "wide" range or much of realistic fields of Physical interest that are known to exist and be manifested in nature.

## 1. Introduction

The standard way to hamiltonize a gauge singular field theory is to follow the Dirac's "multi-generational" algorithm [1,2,3,4]:

In singular field theories, the conjugate momenta are not all independent and therefore, they satisfy one or more constraint equations of the form

$$\Phi(q,p) \approx 0. \qquad (1)$$

These equations are direct result of using the singular Lagrangian, at hand, to define the momenta as functions of coordinates and velocities. The momenta's defining functions have to be satisfied at all times; hence, Eqs.(1) have to be satisfied at all times too, which leads to the consistency conditions

$$\dot{\Phi}(q,p) \approx 0. \qquad (2)$$

These new conditions will lead to one or more of the following [6]: (i) Inconsistency; i.e., the Lagrangian is inconsistent. (ii) Eqs.(2) are identically satisfied using Eqs.(1). (iii) Imposing conditions on the velocities which are not yet explicitly expressible in terms of coordinates, momenta, and other un-expressible velocities only. (iv) New equations of the form (1) which leads to new conditions of the form (2), and so on. Eventually, we reach a stage or "generation" at which no new equations, of the form (1), can be produced.

Constraints, we get at the first stage of this algorithm, are called "primary constraints", and the ones we get at all subsequent stages are called "secondary constraints".

If, by the final stage, we are able to express all the velocities in the theory as functions of the coordinates and the momenta alone, then, this will mean that all the relations of the form (1), which we obtained at all stages, are, exclusively, "second-class constraints", and the theory will not have a "gauge" degeneracy. If, on the other hand, we are not able, by the final stage, to express all the velocities as functions of the coordinates and that momenta alone, then, this indicates we have one or more "first-class constraints": that have vanishing Poisson brackets with all other constraints; which requires the introduction of gauge-fixing conditions that will remove the "gauge" degeneracy caused by the first-class constraints.

This was a brief review of the standard Dirac's algorithm for finding first-class and second-class constraints of different generations: primary and secondary. In Sec.2. we will introduce the proposed conjecture for deriving all the first-class constraints (responsible for the gauge degeneracy), of all stages, in a single step based on the generators of the gauge degeneracy in the Lagrange's formulation, provided that these first-class constraints only involve "canonical" variables, i.e., variables that do not violate the canonical Poisson bracket relations: $\{q_a(\mathbf{x},t), p_b(\mathbf{y},t)\}_{PB} = \delta_{ab}\delta(\mathbf{x}-\mathbf{y})$. In Sec.3., in support of the proposed conjecture, we verify it for many of the realistic fields manifested in nature such as, the free electromagnetic field, **3.1.**, which is the simplest special case of the more general case of Yang-Mills fields, (that describe all realistic fields of nature except gravity), of gauge group *G* interacting with spinor and scalar fields (which we treat in **3.2.**); we show that the first-class constraints that will be produced here match those we get using Dirac's standard method as given in [2,3,4]. Next, in **3.3.**, we verify this conjecture in the case of gravitational field and find the condition on gauge fixing choice for which the first class constraints derived using the Dirac's method matches those derived using the proposed conjecture.

## 2. The Proposed Conjecture

Consider a gauge field theory in the Lagrange's formalism. Its Euler-Lagrange equations must satisfy some gauge identities. These identities can be written in the form [4],[5]:

$$\int \frac{\delta S}{\delta q^a(x)} \Re^a_\alpha(x,x') d^4x \equiv 0, \qquad [\alpha] = r, \qquad (3)$$

where $r$ is the degree of gauge degeneracy, i.e. the total number of independent identities of the form (3). The quantities $\mathfrak{R}_\alpha^a(x,x')$, are called, "gauge generators", and they possess the following properties [5]: (i) They are local functions of space-time, i.e., can be written as

$$\mathfrak{R}_\alpha^a(x,x') = \sum_{k_0=0} \ldots \sum_{k_3=0} \Omega_{\alpha,k_0\ldots k_3}^a(q(x),\partial^\mu q(x),\ldots) \times \partial_0^{k_0}\ldots\partial_3^{k_3}\delta(x-x'), \tag{4}$$

and, (ii) The gauge generators, $\mathfrak{R}_\alpha^a(x,x')$, are independent; i.e., the equations

$$\int \mathfrak{R}_\alpha^a(x,x') f^\alpha(x,q(x),\partial^\mu q(x),\ldots) d^4x = 0,$$

have, $f^\alpha's = 0$, as the only solution.

We know that, in the Hamiltonian formalism, the first-class constraints, we gather from all stages of Dirac's algorithm, are associated with the theory's gauge degeneracy and therefore, equal in number to the degree of gauge arbitrariness. **Now, we propose that**, all the first-class constraints, in a given "realistic" gauge theory, can be written in a form derived from the assumption that, the conjugate momenta, $\pi(x)$'s, inherit the same symmetry (identities) satisfied by the Lagrange's equations, as given by Eqs.(3). This conjecture is restricted to those "canonical" variables that do not violate the canonical Poisson bracket relations between conjugate variables. So, we claim that all first-class constraints that involve "canonical" variables are given by the equations

$$\int \pi_a(x)\, \mathfrak{R}_\alpha^a(x,x')\, d^4x \approx 0, \tag{5}$$

where, the gauge generators $\mathfrak{R}_\alpha^a$ that we use here in (5), are the same ones we have in (3).

## 3. Examples

### 3.1. Free Electromagnetic Field

We start with the simple example of a free electromagnetic field, $A^\mu(x)$, described by the Maxwell Lagrangian, $\mathcal{L} = -\frac{1}{4} F^{\mu\nu} F_{\mu\nu}$. We note first that,

$$\pi_\mu(x) = \frac{\partial \mathcal{L}}{\partial \dot{A}^\mu} = F_{0,\mu} - F_{\mu,0},$$

and in particular,

$$\pi_0(x) = \frac{\partial \mathcal{L}}{\partial \dot{A}^0} \equiv 0. \tag{6}$$

So $\pi_0(x)$, (unlike $\pi_i$'s), is incompatible with canonical Poisson bracket relations and hence, will not be derived (as first-class constraint) using (5).

The Euler-Lagrange equations are

$$\frac{\delta S}{\delta A^{\nu}(x)} \equiv \partial^{\mu} F_{\mu\nu}(x) = 0 \ . \tag{7}$$

They satisfy the identity:

$$\partial^{\nu} \frac{\delta S}{\delta A^{\nu}(x)} \equiv \partial^{\nu}\partial^{\mu} F_{\mu\nu}(x) \equiv 0, \tag{8}$$

which we put in the form of (3) as

$$\int \frac{\delta S}{\delta A^{\mu}(x)} \Re^{\mu}(x,x') \, d^4x = -\int \frac{\delta S}{\delta A^{\mu}(x)} \partial^{\mu} \delta(x-x') \, d^4x \equiv 0 \ , \tag{8'}$$

or,
$$\Re^{\mu}(x,x') = -\partial^{\mu}\delta(x-x') \ . \tag{8''}$$

Now, if we substitute $\Re^{\mu}(x,x')$ into (5), we get

$$\partial^{\mu} \pi_{\mu}(x) \approx 0 \ . \tag{9}$$

Due to (6) and following Dirac [2], we discard $A^0(x)$ and $\pi_0(x)$ degree of freedom; so (9) reduces to:

$$\partial^i \pi_i(x) \approx 0 \ . \tag{9a}$$

Eqs.(6) and (9a) are the same two first-class constraints which we always get for the free electromagnetic field using the standard Dirac's algorithm [2,3,4].

## 3.2. Yang-Mills Fields

We now extend the above simple special case of free electromagnetic field to the more general case of Yang-Mills fields associated with non-Abelian groups of gauge transformations and interacting with spinor and scalar fields. [Here, we will follow Ref.[4] in giving the summary of the model and use the notation and the results of applying Dirac's standard method therein]:

The model, at hand, describes the Yang-Mills fields $A^{a\mu}, (a=1,...,r)$, interacting with spinor fields $\psi^{\alpha}, \bar{\psi}^{\beta}$ and scalar fields $\varphi^m, \varphi^{+n}$. The gauge group, $G$, is a product of semi-simple compact group with several $U(1)$'s; $G$ is generated by Hermitian generators: $\Gamma_a, a=1,...,r$, where, for $\xi \equiv (\psi, \bar{\psi}, \varphi, \varphi^+)$, we have: $\xi(x) \xrightarrow{g \in G} \xi'(x) = \exp(i v^a \Gamma_a)\xi(x)$, with $\Gamma_a$ has the representation:

$$\Gamma_a = \begin{pmatrix} T_a & & & \\ & \bar{T}_a & & \\ & & \tau_a & \\ & & & \bar{\tau}_a \end{pmatrix}, \qquad \text{where } T_a = -\gamma^0 \bar{T}_a \gamma^0 \text{ and } \tau_a = -\bar{\tau}_a^*.$$

$\Gamma_a$'s satisfy the Lie algebra, $[\Gamma_a, \Gamma_b] = i f_{ab}^{\ c} \Gamma_c$.

The Lagrangian of the model (that is gauge-invariant under $G$) is of the form [4]

$$\mathcal{L} = -\frac{1}{4} G_{\mu\nu}^a G^{a\mu\nu} + i \bar{\psi}^\alpha \gamma^\mu \nabla_{\mu\beta}^\alpha \psi^\beta + \bar{\nabla}_{\mu n}^h \varphi^{+n} \nabla_m^{\mu h} \varphi^m - M(\psi, \bar{\psi}, \varphi, \varphi^+), \tag{10}$$

where
$$G_{\mu\nu}^a = \partial_\mu A_\nu^a - \partial_\nu A_\mu^a + f_{bc}^{\ a} A_\mu^b A_\nu^c,$$
$$\nabla_{\mu\beta}^\alpha = \partial_\mu \delta_\beta^\alpha - i T_{a\beta}^{\ \alpha} A_\mu^a,$$
$$\nabla_{\mu m}^h = \partial_\mu \delta_m^h - i \tau_{am}^{\ h} A_\mu^a,$$
$$\bar{\nabla}_{\mu n}^h = \partial_\mu \delta_n^h - i \bar{\tau}_{an}^{\ h} A_\mu^a.$$

In order for $\mathcal{L}$ (given by (10)) to be gauge-invariant: $M(\psi, \bar{\psi}, \varphi, \varphi^+)$ should satisfy the condition

$$\frac{\partial M}{\partial \xi^k} (\Gamma_a)_{k'}^k \xi^{k'} = 0.$$

The Euler-Lagrange equations of this model are:

$$\frac{\delta S}{\delta A_\mu^a} = \mathcal{D}_{vb}^a G^{v\mu b} + \bar{\psi}^\alpha \gamma^\mu T_{a\beta}^{\ \alpha} \psi^\beta - i \bar{\tau}_{an}^{\ k} \varphi^{+n} \nabla_m^{\mu k} \varphi^m - i \bar{\nabla}_n^{\mu k} \varphi^{+n} \tau_{am}^{\ k} \varphi^m = 0,$$

$$\frac{\delta S}{\delta \psi^\alpha} = -i \bar{\psi}^\beta \gamma^\mu \left( \overleftarrow{\partial}_\mu \delta_\alpha^\beta + i T_{a\alpha}^{\ \beta} A_\mu^a \right) - \frac{\partial M}{\partial \psi^\alpha} = 0,$$

$$\frac{\delta S}{\delta \bar{\psi}^\beta} = -i \gamma^\mu \nabla_{\mu\alpha}^\beta \bar{\psi}^\alpha - \frac{\partial M}{\partial \bar{\psi}^\beta} = 0,$$

$$\frac{\delta S}{\delta \varphi^m} = -\nabla_{\mu k}^m \nabla_n^{\mu k} \varphi^{+n} - \frac{\partial M}{\partial \varphi^m} = 0,$$

$$\frac{\delta S}{\delta \varphi^{+n}} = -\nabla_{\mu k}^n \nabla_m^{\mu k} \varphi^m - \frac{\partial M}{\partial \varphi^{+n}} = 0,$$

$$\tag{11}$$

(where, $\mathcal{D}_{vb}^a = \partial_v \delta_b^a + f_{cb}^{\ a} A_v^c$).

Since, $\pi_{A_0^a} = 0$, $(\forall a)$ at all times; and following Dirac [2], we discard the degrees of freedom of $A_0^a$'s. Discarding $A_0^a$'s and $\pi_{A_0^a}$'s, and using (11) we arrive at the identities [4]:

$$\mathcal{D}_{kb}^a \frac{\delta S}{\delta A_k^b} - i \left( \frac{\delta S}{\delta \psi^\alpha} T_{a\alpha'}^{\ \alpha} \psi^{\alpha'} + \frac{\delta S}{\delta \bar{\psi}^\beta} \bar{T}_{a\beta'}^{\ \beta} \bar{\psi}^{\beta'} + \frac{\delta S}{\delta \varphi^m} \tau_{am'}^{\ m} \varphi^{m'} + \frac{\delta S}{\delta \varphi^{+n}} \bar{\tau}_{an'}^{\ n} \varphi^{+n'} \right) \equiv 0, \tag{12}$$

where $k = 1, 2, 3$.

Identities (12) can be put in the form (3), (i.e., $\int \frac{\delta S}{\delta q^w(x)} \Re_a^w(x,y)dx \equiv 0$), if for $q \equiv (A,\xi) \equiv (A,\psi,\bar{\psi},\varphi,\varphi^+)$, we identify:

$$\Re_a^w(x,y) \equiv \begin{cases} \mathcal{D}_{ka}^b \delta(x-y), & w = \binom{b}{k}; k=1,2,3, \\ iT_{a\alpha}^\alpha \psi^{\alpha'}(x)\delta(x-y), & w = \alpha, \\ i\bar{T}_{a\beta}^\beta \bar{\psi}^{\beta'}(x)\delta(x-y), & w = \beta, \\ i\tau_{am}^m \varphi^{m'}(x)\delta(x-y), & w = m, \\ i\bar{\tau}_{an}^n \varphi^{+n'}(x)\delta(x-y), & w = n. \end{cases}$$

(13)

Finally; substituting (13) into the conjectural-rule of (5): we recover the first-class constraints given in [4] (using the standard Dirac's algorithm); namely we get:

$$\mathcal{D}_{kb}^a \pi_{A^{bk}} + i\pi_{\xi^t} (\Gamma_a)_{t'}^t \xi^{t'} \approx 0; \quad k = 1,2,3.$$

As for the first-class constraint, $\pi_{A_0^a} \approx 0$, we are not going to reproduce it using (5), since this constraint is incompatible with canonical Poisson brackets.

### 3.3. Gravitational Field

Classically, and in the absence of matter fields, the Einstein action corresponding to the (symmetric) gravitational field, $g^{\mu\nu}(x)$, is given by [4,7]

$$S_E = \int \mathcal{L}_E d^4x, \quad (\text{with}, \quad \mathcal{L}_E = -\sqrt{-g}R), \tag{14}$$

where, $g \equiv \det\|g_{\mu\nu}\|$; $R$ is the scalar curvature: $R \equiv g^{\mu\nu}(x)R_{\mu\nu}$; and $R_{\mu\nu}$ is the Ricci tensor defined in terms of the Christoffel symbols, $\Gamma_{\mu\nu}^\lambda$, as:

$$R_{\mu\nu} = \Gamma_{\mu\nu,\lambda}^\lambda - \Gamma_{\mu\lambda,\nu}^\lambda + \Gamma_{\mu\nu}^\lambda \Gamma_{\lambda\sigma}^\sigma - \Gamma_{\mu\sigma}^\lambda \Gamma_{\nu\lambda}^\sigma,$$

where, $\Gamma_{\mu\nu}^\lambda = \frac{1}{2} g^{\lambda\sigma} (g_{\sigma\mu,\nu} + g_{\sigma\nu,\mu} - g_{\mu\nu,\sigma})$.

Varying the Einstein action with respect to the gravitational field, we arrive at the Einstein equations of motion:

$$\frac{\delta S_E}{\delta g_{\mu\nu}} \equiv -\sqrt{-g}\left(R^{\mu\nu} - \frac{1}{2}g^{\mu\nu}R\right) = 0. \tag{15}$$

The Einstein action is invariant under general coordinate transformation (general covariance); and the variation of the metric field caused by a gauge transformation due to an infinitesimal coordinate transformation ($x' = x + \delta\xi$), (keeping only terms up to linear in coordinate variation), is given by [4]:

$$\delta g_{\mu\nu} = g_{\mu\nu,\lambda}\delta\xi^\lambda + g_{\mu\lambda}\delta\xi^\lambda_{,\nu} + g_{\lambda\nu}\delta\xi^\lambda_{,\mu}. \tag{16}$$

Equation (16) can be written in terms of the generators of the gauge transformations as

$$\delta g_{\mu\nu}(x) = \int R_{\mu\nu|\lambda}(x,y)\,\delta\xi^\lambda(y)\,dy,$$

where the generators are

$$R_{\mu\nu|\lambda}(x,y) = [g_{\mu\nu,\lambda}(x) + g_{\mu\lambda}(x)\partial_\nu + g_{\lambda\nu}(x)\partial_\mu]\delta(x-y). \tag{17}$$

These gauge generators, (17), are the generators of the "four" identities among the ten Einstein equations of motion, (associated with each of the ten symmetric metric tensor fields), given by (15): Since Einstein equations are "generally covariant", i.e., the equations of motion are invariant under any element of the group of continuous differentiable non-singular coordinate transformation; then four identities, (equal in number to the dimension of the transformation group), should exist. These identities are equal to the "contracted Bianchi identities":

$$-D_\mu\left(\frac{1}{\sqrt{-g}}\frac{\delta S_E}{\delta g_{\mu\nu}}\right) = D_\mu\left(R^{\mu\nu} - \frac{1}{2}g^{\mu\nu}R\right) \equiv 0, \tag{18}$$

Where, $D_\mu$ is the covariant derivative. Expanding the left hand side of (18); we can write these identities in terms of the gauge generators, $R_{\mu\nu|\lambda}(x,y)$, [given by (17)]:

$$\int \frac{\delta S_E}{\delta g_{\mu\nu}(y)} R_{\mu\nu|\lambda}(y,x)\,dy \equiv 0. \tag{19}$$

Since, $p^{\mu 0}(x) \equiv \dfrac{\partial \mathcal{L}_E(x)}{\partial g^{\mu 0}(x)} = 0$, [where $p^{\mu\nu}$ is the (symmetric) momentum conjugate to $g^{\mu\nu}$], and similar to what we did in the two examples above (i.e., **3.1** and **3.2**) when the canonical Poisson relations were violated; we discard the degree of freedom, $h_{0\mu}(=h_{\mu 0})$, by setting it equal to zero. (Where $h_{\mu\nu}$ is the deviation of $g_{\mu\nu}$ from the flat metric, i.e., $g_{\mu\nu} = h_{\mu\nu} + \eta_{\mu\nu}$).

Now we are in position to use [Eq.(5)] along with Eqs.(17,19) to derive the first-class constraints (involving, only, variables compatible with the canonical Poisson relations). These first-class constraints are:

$$\Omega_\lambda(x) \equiv \int p^{ij}(y) R_{ij|\lambda}(y,x)\, dy \approx 0, \qquad (20)$$

where, $i, j = 1, 2, 3$.

Using the gauge generators expressions as given in (17); and define, $\gamma_{ij}^k$, as the Christoffel symbols for the three-dimensional metric, $g^{ij}$; we get:

$$\begin{aligned}\Omega^k(x) &\equiv g^{kl}(x) \int p^{ij}(y) R_{ij|l}(y,x)\, dy \\ &= g^{kl}\left[g_{ij,l} p^{ij} - \partial_j(g_{li} p^{ij}) - \partial_i(g_{lj} p^{ij})\right] \\ &= -2\left(\partial_i p^{ik} + \gamma_{ij}^k p^{ij}\right),\end{aligned}$$

(21a)

where, $k = 1, 2, 3$.

We also get:

$$\begin{aligned}\Omega^0(x) &\equiv g^{00}(x) \int p^{ij}(y) R_{ij|0}(y,x)\, dy \\ &= g_{ij,0} p^{ij} - \partial_j(g_{0i} p^{ij}) - \partial_i(g_{0j} p^{ij}) \\ &= g_{ij,0} p^{ij},\end{aligned}$$

(21b)

Where we used that (as indicated above): $h_{0\mu}(= h_{\mu 0}) = 0$; [i.e. $g_{\mu 0}(x) = \eta_{\mu 0}(x)$; where the flat metric has signature= -2].

Our results in (21a), (21.b) can now be compared with the results of Ref.[4], where Dirac algorithm was carried out and applied to the Fadeev-Lagrangian [8], (with the metric fields are taken to reduce to the asymptotically flat ones at infinity). Fadeev-action differs from Einstein-action, only, by a surface term; and hence both give the same equations of motion that satisfy the same gauge identities, namely, the contracted Bianchi identities, (18).

There are "four" first-class constraints, $T^\mu$'s, [given in Ref.[4]: Eq.(4.4.18)], that are left after discarding the $h_{0\mu}$'s (and $p_{0\mu}$'s) degrees of freedom.

$T^i$'s match $\Omega^i$'s, given by (21a), namely:

$$\Omega^k = -2\left(\partial_i p^{ik} + \gamma_{ij}^k p^{ij}\right) = T^k, \qquad k = 1, 2, 3. \qquad (22)$$

Next, we simplify $T^0$; using Expressions for $z_{ik}$ and $p^{ik}$ as given in [4]:

$$z_{ik} = g_{ik,0} - g_{0k,i} - g_{0i,k} + 2\gamma_{ik}^l g_{0l};$$

$$p^{ik} \equiv \frac{\partial \mathcal{L}}{\partial h_{ik,0}} = \frac{1}{2}\sqrt{-g^{00} g_{(3)}}\left(e^{il} e^{km} - e^{ik} e^{lm}\right) z_{lm},$$

(23)

where, $g_{(3)} = \det \|g_{ik}\|$; $e^{ik} g_{kj} = \delta^i_j$.

Since, (as indicated above), $g_{\mu 0}(x) = \eta_{\mu 0}(x)$; then (23) reduces to

$$z_{ik} = g_{ik,0} \quad ; \quad p^{ik} = \frac{1}{2}\sqrt{-g_{(3)}} \left( e^{il} e^{km} - e^{ik} e^{lm} \right) z_{lm}. \quad (24)$$

Using (24); $T^0$ is given for Fadeev-Lagrangian [4,8] as,

$$\begin{aligned} T^0 &= \sqrt{-\frac{1}{g_{(3)}}} p^{ik} \left( g_{il} g_{km} - \frac{1}{2} g_{ik} g_{lm} \right) p^{lm} + \sqrt{-g_{(3)}} R_{(3)} \\ &= \frac{1}{2} p^{kl} z_{kl} + \sqrt{-g_{(3)}} R_{(3)} \\ &= \frac{1}{2} p^{kl} g_{kl,0} + \sqrt{-g_{(3)}} R_{(3)}, \end{aligned}$$

(25)

where, $R_{(3)}$ is the 3-curvature scalar.

Eqs. (21b) and (25) show that both $\Omega^0(x)$ and $T^0(x)$ define the same first-class constraint provided that $\sqrt{-g_{(3)}} R_{(3)}$ vanishes on the constraint surface on account of an appropriate choice of gauge-fixing supplementary conditions.

## 4. Conclusion

It was shown above that the proposed conjectural-rule, given by (5), does work in reproducing the first-class constraints (that don't involve variables that violate the canonical Poisson brackets), for much of the known realistic fields. All first-class constraints, (involving only variables compatible with canonical Poisson brackets), of the free electro-magnetic field, the interacting Yang-Mills fields and the gravitational field were recovered and matched those derived using the standard Dirac's multi-generational algorithm. The examples of Sec.3 show that (5) will hold true for much of the fields of physical interest. Definitely, we make here no claim that conjecture (5) holds for any arbitrarily constructed Lagrangian; but nevertheless, this conjecture seems to work strikingly well for "realistic" fields that exist in nature.

## 5. Acknowledgment

I thank the Ilfat & Bah. Foundation (ed'Oreen, Btouratij) for its continuous support. I thank Mrs and Mr Shahrouj for useful discussions.